\documentclass[twocolumn,showpacs,amsmath,amssymb]{revtex4}
\usepackage[dvips]{graphics}
\def\imo{i}
\def\re#1{Re\left(#1\right)}
\def\im#1{Im\left(#1\right)}

\begin{document}
\title{Stability of tardyons and tachyons in the rotating and expanding Universe}
\author{R. A. Konoplya}\email{konoplya_roma@yahoo.com}
\affiliation{\mbox{DAMTP, Centre for Mathematical Sciences, University of Cambridge,}\\ Wilberforce Road, Cambridge CB3 0WA, UK.}
\author{A. Zhidenko}\email{olexandr.zhydenko@ufabc.edu.br}
\affiliation{Centro de Matem\'atica, Computa\c{c}\~ao e Cogni\c{c}\~ao, Universidade Federal do ABC (UFABC),\\ Rua Santa Ad\'elia, 166, 09210-170, Santo Andr\'e, SP, Brazil}
\begin{abstract}
  In the present paper we analyze the spectrum of quasinormal modes for massive scalar and Dirac fields, allowing for both tardyonic ($\mu^2 >0$) and tachyonic ($\mu^2 <0$) masses, in the expanding and rotating cosmological background. The spectrum found shows a number of peculiar features, which are absent in the Minkowski space-time. A hypothetical particle that moves faster than light, \emph{a tachyon}, is known to be classically unstable in the Minkowski space-time. This instability has its analog at the quantum level: small vacuum fluctuations of the field lead to the unbounded growth of the amplitude, so that the appearance of the real tachyons in the spectrum means that there is catastrophic instability in the theory. It has been conjectured a long time ago that possibly the lightest particles with nonzero mass, the neutrino, may be a tachyon. Here we shall show that in the rotating and expanding Universe tachyons are stable if their mass is less than some constant, which is related to the Universe's rotation and expansion scales. The current upper bound on the rotation scale gives us a very small upper bound on the tachyon's mass which is many orders less than the mass of the electron. This might explain why only very light particles have the chance of becoming tachyons. It is shown that the spectrum of the ``normal'' ($\mu^2 >0$) Dirac field has a discontinuity as a function of the z-component of the wave vector $k_z$ at $k_z = 0$.
\end{abstract}
\maketitle

\section{Introduction}

The spectrum of classical fields in various cosmological space-times is a well-established area, which is a starting point for quantization of the fields in the curved backgrounds. By now it includes a detailed knowledge of spectra for \emph{massless} scalar and Dirac fields \cite{Leahy:1982dj,Hiscock:1978iq,GuhaThakurta:1980ax} in the rotating (but not expanding) G\"odel-like universes \cite{Godel:1949ga,Tiomno}. The massless scalar field in the expanding and rotating background was considered by Panov \cite{Panov}. Yet, all these works were devoted solely to ``normal'' nontachyonic fields.

The first comprehensive study of would be particles moving faster than light, i.e. tachyons, was done as early as 1962 \cite{Bilaniuk}. Since that time there have been a number of other theoretical studies on the various properties of these hypothetical particles in quantum field theory, string theory, and gravitation \cite{tachyons}.
One of the main problems with tachyons is their instability. Indeed, one can easily see that in the Minkowski space-time $ds^2= dt- d x^2 - d y^2 - d z^2$, the Klein-Gordon equation with the negative values of $\mu^2$
\begin{equation}\label{KGequation}
\phi^{;\mu}_{;\mu}+\mu^2=\frac{1}{\sqrt{-g}}\left(\sqrt{-g}g^{\mu\nu}\phi_{,\mu}\right)_{,\nu}+\mu^2\phi=0
\end{equation}
allows for the normalizable exponentially growing modes $\phi \propto e^{-\imo \omega t}$, where $\omega = \re{\omega} + \imo\im{\omega}, ~ \im{\omega} > 0$.
This means that vacuum fluctuations of the field will grow without the bound leading to the instability and decay of the tachyonic state.
Strictly speaking, the above instability creates unsurpassable difficulties in quantization, so that the ``tachyon'' has never been considered seriously by the majority of researchers as a real quantum particle, but rather like a synonym of the onset of instability in the system.
Tachyons also play an important role in cosmology, where they are used in the spontaneous symmetry breaking \cite{Linde:1982uu}.


Theoretically, the tachyonic nature of the neutrino was proposed in the 1980s \cite{Chodosa}, and once one accepts the conception of the tachyon, it would be natural to expect a number of different tachyonic particles with various masses.
Yet, today we observe only single particles, neutrino, which potentially could be a tachyon. Though, there are also arguments against the tachyonic nature of the neutrino, to which we shall possibly contribute one more argument here.
If one supposes the tachyonic nature of a superluminal neutrino, then a straightforward question is why we observe only a single candidate for the tachyon, with a mass that is apparently much smaller than the electron mass and why ``heavy'' tachyons cannot be observed?
In this work we shall show that only very light tachyons can be classically stable in a number of rotating cosmological backgrounds.
At first glance, this could answer both questions posed above: the instability and the selectivity in favor of the lightest particles. The Universe is supposed to be expanding, so that the initial rotation should have mostly been damped today.
Therefore, as only a very slow rotation of the Universe is allowed by current observations, the stability is possible only for those tachyons that are many orders lighter than the electron. Yet, we observed that the threshold of instability occurs at a tiny mass which seems to be too small to provide superluminality of the Standard Model particles.

Previously, one of us reported on the stability gap for tachyons in the pure G\"odel-like nonexpanding space-times \cite{Konoplya:2011ag}.
Here we shall generalize this result to a more realistic situation of expanding, rotating Universe and shall make estimates for the upper bound of tachyon's mass from the current upper bound on the rotation rate of the Universe and the Hubble constant.

Another aim of our research, which is not connected to supeluminality issues, is analysis of spectra of ``normal'' (tardyon) fields in the expanding and rotating Universe.
We have found a number of distinctions from the previously studied cases of scalar fields in these backgrounds, which may raise new problems under quantization of these fields.

The work is organized as follows. Sec.~\ref{sec:scalarfield} is devoted to the stability of a scalar tachyonic field, and Sec.~\ref{sec:Diracfield} deals with massive Dirac tachyons. Sec.~\ref{sec:QNMs} summarizes the spectrum of tardyonic scalar and Dirac fields. In Sec.~\ref{sec:conclusions} we discuss the obtained results and open problems.

\section{Scalar field}\label{sec:scalarfield}

A wide class of rotating and expanding cosmological solutions can be described by the G\"odel-like metric
\begin{equation}\label{metric}
ds^2=\left(dt+a(t)e^{\alpha x}dy\right)^2-a(t)^2\left(dx^2+B^2e^{2\alpha x}dy^2+dz^2\right),
\end{equation}
which includes both causal and noncausal generalizations \cite{Vaidya:1967zz,Ozsvath,Tiomno,Barrow:1998wa} of the classical G\"odel solution \cite{Godel:1949ga}. The G\"odel metric is reproduced if $B^2 = 1/2$ and $a(t)=1$.

Although here we shall use the Cartesian coordinates, let us note that the metric (\ref{metric}) can be rewritten in the cylindrical coordinates $(t', r, \phi, z)$ as follows:
\begin{eqnarray}
 d s^2=\left(2a(t')\sinh^2\left(\frac{\alpha r}{2}\right)\frac{d \phi}{B\alpha} + d t' \right)^2 - \\\nonumber
 a^2(t')\left(d r^2 + d z^2 + \frac{\sinh^2(\alpha r)}{\alpha^2} d \phi^2\right).
\end{eqnarray}
If the following inequality is fulfilled
\begin{equation}
1+ \frac{B^2-1}{B^2}\sinh^2\left(\frac{\alpha r}{2}\right) >0,
\end{equation}
or simply,
\begin{equation}
B^2 \geq 1,
\end{equation}
no closed time-like curves are allowed for the space-time. This is an example of the causal cosmological solutions suggested in \cite{Tiomno}. One can see that value of $B$ does not change our estimations for the order of the tachyon mass, so that, qualitatively, our conclusions will not depend on the causality of the background space-time.

Following \cite{Panov} we introduce the new coordinates $\tau$ and $\eta$ instead of $t$ and $x$,
$$d\tau=\frac{dt}{a(t)}, \qquad \eta=e^{-\alpha x}.$$
In terms of these new coordinates the metric takes form,
\begin{eqnarray}\nonumber
&\displaystyle ds^2=a(\tau)^2\left(\left(d\tau+\frac{dy}{\eta}\right)^2-\frac{d\eta^2}{\alpha^2\eta^2}-\frac{B^2dy^2}{\eta^2}-dz^2\right),&\\
&\displaystyle g=-\frac{B^2a^8(\tau)}{\alpha^2 \eta^4}.&
\end{eqnarray}
We can separate the variables in the Klein-Gordon equation (\ref{KGequation}), using the following ansatz
\begin{equation}
\phi(\tau,\eta,y,z)=e^{\imo k_yy+\imo k_zz}R(\tau,\eta).
\end{equation}
Then, the wave-like equation can be written as follows:
\begin{eqnarray}
\displaystyle\frac{\partial^2R}{\partial \eta^2}+\frac{1-B^2}{B^2\alpha^2\eta^2}\left(\frac{\partial^2R}{\partial \tau^2}+\frac{a'(\tau)}{a(\tau)}\frac{\partial R}{\partial \tau}\right) -\frac{2\imo k_y}{B^2\alpha^2\eta}\frac{\partial R}{\partial \tau}\\\nonumber
\displaystyle-\left(\frac{k_y^2}{B^2}+\frac{k_z^2+a^2(\tau)\mu^2}{\eta^2}\right)\frac{R}{\alpha^2}-\frac{a'(\tau)}{a(\tau)}\frac{2\imo k_y R}{B^2\alpha^2\eta}=0.
\end{eqnarray}
Let us be limited by an adiabatic expansion of the universe, i.e. we shall consider the radius of the universe $a(\tau)$ and the Hubble parameter $H$ as constants:
\begin{equation}
a(\tau)=a,\qquad \frac{1}{a(t)}\frac{da}{dt}=\frac{a'(\tau)}{a^2(\tau)}=H.
\end{equation}
Then, following \cite{Panov}, we introduce the frequencies of the proper oscillations of the universe as
\begin{equation}
R(\tau,\eta)=\sum_{\omega}e^{-\imo\omega\tau}N_\omega(\eta),
\end{equation}
where $N_{\omega}(\eta)$ satisfies the following equation
\begin{eqnarray}\label{scalar-wavelike}
N_\omega''(\eta)-\left(\frac{k_y^2}{B^2}+\frac{2k_y\omega+2\imo k_y aH}{B^2\eta} + \frac{k_z^2+a^2\mu^2}{\eta^2}\right. \\\nonumber \left.+\frac{(1-B^2)(\omega^2+\imo\omega aH)}{B^2\eta^2}\right) \frac{N_\omega(\eta)}{\alpha^2}=0.
\end{eqnarray}

Eq. (\ref{scalar-wavelike}) has a regular singular point at $\eta=0$ and an irregular singular point at $\eta=\infty$ with the asymptotical behavior
$$N_{\omega}\propto \exp\left(\pm \frac{k_y\eta}{B\alpha}\right).$$

Let us notice, that the asymptotically G\"odel space-time, as well as the asymptotically anti-de Sitter (AdS) one, is not globally hyperbolic, so that mathematically strict analytical proof of the stability should also include the detailed consideration of the well-posedness of the initial value problem. We shall study only normalizable solutions, i.e.
\begin{equation}\label{norma}
\intop_0^{\infty}|\phi|^2\sqrt{-g}~d\eta<\infty\quad\Longrightarrow\quad\intop_0^{\infty}\frac{|N_{\omega}(\eta)|^2}{\eta^2}d\eta<\infty.
\end{equation}
In order to satisfy convergence conditions at both endpoints of the integral we require
\begin{eqnarray}\label{zerobc}
&&N_{\omega}(\eta\rightarrow0)\propto\eta^p, \qquad p>1/2;\\\label{infbc}
&&N_{\omega}(\eta\rightarrow\infty)\propto \exp\left(-\frac{|k_y|\eta}{B\alpha}\right).
\end{eqnarray}

In terms of the new variable $\xi$,
$$\xi=2\frac{|k_y|\eta}{B\alpha},$$
the equation takes the following form
\begin{equation}
N_\omega''(\xi)-\left(\frac{1}{4}+\frac{J}{\xi}+\frac{K}{\xi^2}\right)N_\omega(\xi)=0,
\end{equation}
where
\begin{eqnarray}
J &=& \frac{m (\omega+\imo aH)}{B\alpha},\\
K &=& \frac{k_z^2+a^2\mu^2}{\alpha^2}+\frac{(1-B^2)(\omega^2+\imo\omega aH)}{\alpha^2B^2}.
\end{eqnarray}
and $ m = sgn(k_y)$.
Taking into account the asymptotical behavior of the wave equation, we introduce a new function $P(\xi)$,
\begin{equation}
N_\omega(\xi)=e^{-\xi/2}\xi^{q}P(\xi).
\end{equation}
Choosing $q$ such that $K=q(q-1)$, we find that the function $P(\xi)$ satisfies the Kummer's equation
\begin{equation}\label{Kummer-like}
\xi P''(\xi)+(2q-\xi)P'(\xi)=(q+J)P(\xi).
\end{equation}
The only solution, which satisfies (\ref{zerobc}) as $\xi\rightarrow0$, is
\begin{equation}
P(\xi)=\left\{
                       \begin{array}{ll}
                         M(q+J,2q,\xi), & \hbox{if $\re{q}>1/2$;} \\
                         \xi^{1-2q} M(J-q+1,2-2q,\xi), & \hbox{if $\re{q}<1/2$;}
                       \end{array}
                     \right.
\end{equation}
where $M(\alpha,\beta,\xi)$ is the confluent hypergeometric function.

As $\xi\rightarrow\infty$, $M(\alpha,\beta,\xi)$ grows exponentially and violates (\ref{infbc}) unless $\alpha$ is a nonpositive integer. Thus,
\begin{equation}\label{fspectreq}
-N=\left\{
                       \begin{array}{ll}
                         q+J, & \hbox{if $\re{q}>1/2$;} \\
                         J-q+1, & \hbox{if $\re{q}<1/2$.}
                       \end{array}
                     \right.
\end{equation}
where $N=0,1,2,3\ldots$ Finally, we find
\begin{equation}\label{spectreq}
K=(N+J)(N+J+1).
\end{equation}
A general solution to (\ref{spectreq}) gives us an analytic expression for the frequency $\omega$:
\begin{widetext}
\begin{equation}\label{square-root1}
\omega=-\frac{(1\!+\!2N)\alpha}{2 m B}\!-\!\imo aH\frac{(1\!+\!B^2)}{2B^2} \pm\sqrt{\frac{\alpha^2\!+\!4(k_z^2\!+\!\mu^2)\!-\!a^2H^2(B^{-2}\!-\!1)^2}{4}+\left(\frac{\imo aH(1\!+\!2N)\alpha}{2B m}+\frac{(1\!+\!2N)^2\alpha^2}{4}\right)(B^{-2}\!-\!1)}.
\end{equation}
\bigskip
\end{widetext}
When $B^2=1/2$, $H=0$ and $\alpha = \sqrt{2} \Omega$, we have the nonexpanding G\"odel space-time. When the expression under the square root is positive in the above equation, $\omega$ is pure real, so that the tachyon is stable. Thus, the upper bound on the tachyon's mass reads
\begin{equation}
|\mu| < \Omega,
\end{equation}
which coincides with Eq.~(13) of \cite{Konoplya:2011ag}.

When $B=1$, the expression for $\omega$ is formally reduced to the one in \cite{Panov}:
$$\omega=-\frac{(1+2N)\alpha\pm\sqrt{\alpha^2+4(k_z^2+a^2\mu^2)}}{2 m}-\imo a H.$$

There are unstable modes in the spectrum if
$$\alpha^2+4a^2\mu^2 > -4a^2H^2,$$
allowing for tachyons with very small mass
\begin{equation}
|\mu|\leq\sqrt{\frac{\alpha^2}{4a^2}+H^2}=H\sqrt{1+\frac{\Omega_H^2}{2}},
\end{equation}
where $\Omega_H=\Omega/H=\alpha/\sqrt{2}aH$ is the rotation parameter measured in units of the Hubble constant. Estimations in \cite{Obukhov:2000jf} show that $\Omega_H$ can be equal to $74$.

The Hubble constant is of order $10^{-10}year^{-1}$, giving an estimation for the tachyon mass
\begin{equation}\label{massest}
|\mu|\lesssim10^{-8}year^{-1}\hbar/c^2\sim 10^{-66}kg\sim 10^{-36}m_e,
\end{equation}
where $m_e$ is the electron mass.

The estimation for the upper limit on the Universe's rotation certainly crucially depends on the cosmological model under consideration (see for instance a recent review on the observational aspects of the global rotation \cite{Godlowski:2011rf}). Therefore, the upper limit for the global rotation that we have used is most physically relevant as it deals with the same cosmological model.

Apparently, the most authoritative approach to finding an upper constraint on the Universe`s rotation is based on the Cosmic Microwave Background Anisotropies (CMBA) data of NASA, owing to their precision. Constraints for nonflat cosmological models, including the G\"odel-Friedman ones, are in the range $10^{-9}\div10^{-8} rad/year$ \cite{Obukhov:2000jf,Godlowski:2011rf,SuChu,Williams:2011jm}. In particular, in \cite{Williams:2011jm} the analysis of the so-called Sachs-Wolf effect (the gravitational redshift of photons in the cosmic microwave background, which makes the cosmic microwave background spectrum uneven) leads to the $10^{-9} rad/year$ constraint for the global rotation. If one considers this constraint of $10^{-9} rad/year$ as physically relevant, then the estimated upper limit on mass stability will be one order less than in Eq.~(\ref{massest}).

\vspace{3mm}
\section{Dirac field}\label{sec:Diracfield}

The Dirac equation in a curved space-time has the form
\begin{equation}
-\imo\gamma^ae_a^{~\mu}(\Phi_{,\mu}+\Gamma_\mu\Phi)+\mu\Phi=0,
\end{equation}
where $\gamma_a$ are Dirac matrices
$$\{\gamma^a\gamma^b\}=2\eta^{ab},$$
and $e_a^{~\mu}$ is the tetrad field which satisfies
$$ g^{\mu\nu}=\eta^{ab}e_a^{~\mu}e_b^{~\nu}, \quad g^{\mu\nu}=\eta^{ab}e_a^{~\mu}e_b^{~\nu}.$$
One can choose the tetrad field as follows
\begin{equation}
e_a^{~\mu}=\frac{1}{B a(\tau)}\left(
             \begin{array}{cccc}
               B & 0 & 0 &0 \\
               0 & B\alpha\eta & 0 & 0 \\
               -1 & 0 & \eta & 0 \\
               0 & 0 & 0 & B \\
             \end{array}
           \right).
\end{equation}

Here we shall use the standard representation for the Dirac matrices
\begin{equation}
\gamma^0=\left(
             \begin{array}{cc}
               0 & \imo \\
               -\imo & 0 \\
             \end{array}
           \right),
\qquad
\gamma^i=\left(
             \begin{array}{cc}
               0 & \imo\sigma^i \\
               \imo\sigma^i & 0 \\
             \end{array}
           \right),
\end{equation}
where $\sigma_i$ are standard Pauli matrices ($i=1,2,3$).

The spin connections $\Gamma_\mu$ are defined as
\begin{equation}
\Gamma_\mu=\frac{1}{8}[\gamma^a,\gamma^b]~g_{\nu\lambda}~e_a^{~\nu}~e_{b~~;\mu}^{~\lambda}.
\end{equation}
For the space-time under consideration it takes the following form
\begin{widetext}
\begin{eqnarray}\nonumber
&\displaystyle\Gamma_0=\frac{\alpha}{8B}\left([\gamma^1,\gamma^2]-\frac{2a'(\tau)}{a(\tau)}[\gamma^0,\gamma^2]\right);\qquad\qquad\qquad
\Gamma_1=\frac{1}{8\eta B}\left([\gamma^0,\gamma^2]+\frac{2a'(\tau)}{\alpha a(\tau)}\left(B[\gamma^0,\gamma^1]+[\gamma^1,\gamma^2]\right)\right);&
\\\nonumber
&\displaystyle\Gamma_2=-\frac{\alpha}{8\eta}\left([\gamma^0,\gamma^1]+\frac{2B^2-1}{B}[\gamma^1,\gamma^2]\right)-\frac{(1-B^2)a'(\tau)}{4\eta Ba(\tau)}[\gamma^0,\gamma^2];\qquad
\Gamma_3=\frac{a'(\tau)}{4B a(\tau)}\left(B[\gamma^0,\gamma^3]-[\gamma^2,\gamma^3]\right).&
\end{eqnarray}
As for the scalar field, we shall consider the adiabatic approximation for the Universe's expansion and assume the following ansatz:
\begin{equation}\label{anzats}
\Phi(\tau,\eta,y,z)=e^{-\imo\omega\tau+\imo k_yy+\imo k_zz}\Psi(\eta),
\end{equation}
Replacing $\Phi$ in the Dirac wave equation by (\ref{anzats}), and after some calculations, one can reduce the wave equations to the following the matrix form:
\begin{eqnarray}
\Psi'(\eta)=\left(
                                     \begin{array}{cc}
                                       P & M \\
                                       M & Q \\
                                     \end{array}
                                   \right)\Psi(\eta), \quad \hbox{where}~~
M&=&\frac{1}{\alpha\eta}\left(
    \begin{array}{cc}
      0 & a\mu\\
      a\mu& 0\\
    \end{array}
  \right), \\\nonumber
P&=&\frac{1}{4B \alpha\eta}\left(
    \begin{array}{cc}
      2(B\alpha - 2 k_y \eta - 3 \imo a H- 2\omega) & -6 B a H - 4 \imo B k_z + 4 \imo B\omega + \imo\alpha \\
      -6 B a H + 4 \imo B k_z + 4 \imo B\omega - \imo\alpha & 2 (B \alpha + 2 k_y \eta + 3 \imo aH + 2 \omega) \\
    \end{array}
  \right),\\\nonumber
Q&=&\frac{1}{4B \alpha\eta}\left(
    \begin{array}{cc}
      2(B\alpha - 2 k_y \eta - 3 \imo a H- 2\omega) & 6 B a H - 4 \imo B k_z - 4 \imo B\omega + \imo\alpha  \\
      6 B aH + 4\imo B k_z - 4\imo B \omega + \imo\alpha & 2 (B\alpha + 2 k_y \eta + 3 \imo aH + 2 \omega) \\
    \end{array}
  \right).
\end{eqnarray}
Choosing the 4-component spinor as
\begin{equation}
\Psi(\eta)=\left(1+\frac{{\tilde k}_z}{a\mu}\gamma^3\right)\left(
                                                               \begin{array}{c}
                                                                 \psi_1(\eta) \\
                                                                 \psi_2(\eta) \\
                                                                 \psi_3(\eta) \\
                                                                 \psi_4(\eta) \\
                                                               \end{array}
                                                             \right),
\qquad {\tilde k}_z = k_z\left(\sqrt{1+\frac{a^2\mu^2}{k_z^2}}-1\right) \qquad ({\tilde k}_z^2=a^2\mu^2, \quad\hbox{if}\quad k_z=0),
\end{equation}
we are able to reduce the matrix equation to the second-order differential equations for each component
\begin{equation}\label{Dirac-wavelike}
\psi''(\eta)-\left(\frac{k_y^2}{B^2}+\frac{2k_y{\tilde\omega}\pm k_y B\alpha}{B^2\eta}+\frac{4F(k_z)^2-\alpha^2}{4\eta^2}+\frac{(1-B^2){\tilde\omega}^2}{B^2\eta^2}\right)\frac{\psi(\eta)}{\alpha^2}=0,
\end{equation}
where ${\tilde\omega}=\omega+3\imo aH/2$. For left-handed components
\begin{equation}\label{left-handed}
F(k_z)=k_z\frac{a^2\mu^2-{\tilde k}_z^2}{a^2\mu^2+{\tilde k}_z^2}+2{\tilde k}_z\frac{a^2\mu^2}{a^2\mu^2+{\tilde k}_z^2}+\frac{\alpha}{4B} \qquad \left(\lim_{\mu\rightarrow0}F(k_z)=k_z+\frac{\alpha}{4B}\right).
\end{equation}
In the limit $H\rightarrow0$, $\mu\rightarrow0$ we reproduce formula (102) of \cite{Leahy:1982dj} (one should take $B^2=\alpha=1/2$).

Again, we study only the normalizable solutions
\begin{equation}\label{normaDirac}
\intop_0^{\infty}\overline{\Phi}\gamma_0\Phi\sqrt{-g}~d\eta<\infty\quad\Longrightarrow\quad\intop_0^{\infty}\frac{|\psi(\eta)|^2}{\eta^2}d\eta<\infty,
\end{equation}
which imply the same requirements for the endpoints as for the scalar field (\ref{zerobc}),(\ref{infbc}).

Eq. (\ref{Dirac-wavelike}) has the same form as (\ref{Kummer-like}) with the coefficients $J$ and $K$ given by
\begin{eqnarray}
J = \frac{m{\tilde\omega}}{B\alpha}\pm\frac{1}{2},\qquad
K = \frac{F^2(k_z)}{\alpha^2}-\frac{1}{4}+\frac{(1-B^2){\tilde\omega}^2}{\alpha^2B^2}.
\end{eqnarray}

From (\ref{spectreq}) one finds
\begin{equation}\label{square-root2}
\omega=-\frac{(2N+1\pm1)\alpha}{2Bm}\pm\frac{\sqrt{(2N+1\pm1)^2(B^{-2}-1)\alpha^2+4F^2(k_z)}}{2}-\frac{3\imo aH}{2}.
\end{equation}
The ``-'' sign for $N=0$ violates the conditions of convergence of the Dirac field norm (see \cite{Leahy:1982dj} for details), so that we are able to write
both solutions as
\begin{equation}
\omega=-\frac{(N+1)\alpha}{Bm}\pm\sqrt{(N+1)^2(B^{-2}-1)\alpha^2+F^2(k_z)}-\frac{3\imo aH}{2}.
\end{equation}
\end{widetext}

For $B=1$ the spectrum of the wave-like equation takes the following simple form
$$\omega=-\frac{(N+1)\alpha}{m}\pm F(k_z)-\frac{3\imo aH}{2}.$$
Then, for $B=1$ the instability condition is
$$\im{F(k_z)}>\frac{3aH}{2}.$$
Since the imaginary part of $F(k_z)$ reaches its maximum for $k_z=0$ and
$$max(\im{F(k_z}))=\im{F(0)}= \im{a\mu},$$
the bound for fermion mass does not depend on the Universe's rotation in this case, and,
\begin{equation}
|\mu|\leq\frac{3H}{2}.
\end{equation}

However, for $B^2\neq1$, and, in particular, for the G\"odel case ($B^2=1/2$), the bound on the tachyon mass depends on the rotation parameter.
The instability condition becomes
\begin{equation}
\im{\sqrt{(N+1)^2(B^{-2}-1)\alpha^2+F^2(k_z)}}>\frac{3aH}{2}.
\end{equation}

From the formula
$$|\im{\sqrt{x}}|=\sqrt{\frac{|x|-\re{x}}{2}}\quad (\re{x}>0),$$
we conclude that $N=0$ corresponds to the most unstable mode, for which we find (for $\mu^2<0$) the following expression:
\begin{eqnarray}\nonumber
\sqrt{\left(\frac{\alpha^2(17-16B^2)}{16B^2}+a^2\mu^2\right)^2-\frac{\alpha^2a^2\mu^2}{4B^2}}\qquad\\\nonumber-
\left(\frac{\alpha^2(17-16B^2)}{16B^2}+a^2\mu^2\right)>\frac{9a^2H^2}{2}.
\end{eqnarray}

Finally we find that the tachyon is stable if
\begin{eqnarray}
|\mu|&\leq&\frac{3H}{2}\sqrt{1+\frac{16(1-B^2)}{1+36B^2a^2H^2/\alpha^2}}\\\nonumber&&=\frac{3H}{2}\sqrt{1+\frac{16(1-B^2)}{1+18B^2/\Omega_H^2}}.
\end{eqnarray}

For the G\"odel-like solution ($B^2=1/2$) and the quick rotation scenario ($\Omega_H\gg1$), we obtain the tachyon mass' bound of half order higher than the Hubble parameter
\begin{equation}
|\mu|\lesssim\frac{9H}{2}.
\end{equation}

Let us notice, that if $B^2>1$ the expression inside the square root in (\ref{square-root1}), (\ref{square-root2}) can always be negative for sufficiently large $N$, even for the massless case.
Thus, for $B^2>1$ tachyons (both scalar and Dirac) are unstable, yet the instability occurs for the normal matter as well, making this case quite exotic. In summary, noncausal space-times with $B^2<1$ and causal with $B^2=1$  allow for the tachyon's stability, while causal space-time with $B^2>1$ is a quite exotic one, because not only tachyonic matter, but also the normal one is unstable.

\vspace{3mm}
\section{Quasinormal modes of non-tachyonic fields}\label{sec:QNMs}

First let us  note, that solutions of (\ref{spectreq}) are not always also the solutions of (\ref{fspectreq}). Because of the requirement for the real part of $q$ (\ref{fspectreq}), one can see that $J$ must always have a negative real part, which implies that the real part of $\omega$ has an opposite to the $m$ sign and always has a minimal value. That is why if one studies both the real and imaginary parts of $\omega$, he must discard solutions that do not satisfy (\ref{fspectreq}).

When $\mu^2\geq0$ we can prove that Eq. (\ref{fspectreq}) is not satisfied for all of the modes of (\ref{square-root1}) for which the sign of the real part of the square root is $-m$. Indeed, \footnote{Hereafter we fix the sign for a square root of a complex value so, that its real part is nonnegative.},

\begin{widetext}
\begin{eqnarray}
\re{m\omega}
\geq-\frac{(1\!+\!2N)\alpha}{2 B}+\re{\sqrt{\frac{-\!a^2H^2(B^{-2}\!-\!1)}{4B^2}+\left(\frac{\imo aH(1\!+\!2N)\alpha}{2B m}+\frac{(1\!+\!2N)^2\alpha^2}{4}\right)(B^{-2}\!-\!1)}}\\\nonumber=-\frac{(1\!+\!2N)\alpha}{2 B}+\re{\frac{(1\!+\!2N)\alpha}{2}+\frac{\imo a H}{2Bm}}\sqrt{B^{-2}\!-\!1}=-\frac{(1\!+\!2N)\alpha}{2}\left(\frac{1}{B}-\sqrt{B^{-2}\!-\!1}\right)\geq-\frac{(1\!+\!2N)\alpha}{2}B,
\end{eqnarray}
from which it follows that
$\re{J}\geq-N-\frac{1}{2}$.

Therefore, for $\mu^2\geq0$ the quasinormal spectrum is
\begin{equation}\label{square-rootf1}
\omega=-\frac{(1\!+\!2N)\alpha}{2 m B}\!-\!\imo aH\frac{(1\!+\!B^2)}{2B^2} -\frac{1}{m}\sqrt{\frac{\alpha^2\!+\!4(k_z^2\!+\!\mu^2)\!-\!a^2H^2(B^{-2}\!-\!1)^2}{4}+\left(\frac{\imo aH(1\!+\!2N)\alpha}{2B m}+\frac{(1\!+\!2N)^2\alpha^2}{4}\right)(B^{-2}\!-\!1)}.
\end{equation}
Similarly, we can prove that for the Dirac field
\begin{equation}\label{square-rootf2}
\omega=-\frac{(N+1)\alpha}{Bm}-\frac{\sqrt{(N+1)^2(B^{-2}-1)\alpha^2+F^2(k_z)}}{m}-\frac{3\imo aH}{2}, \qquad \mu^2\geq0.
\end{equation}
This was also shown for a particular value of $B$, $B^2=\frac{1}{2}$, in \cite{Leahy:1982dj}.
\end{widetext}

\begin{figure*}
\resizebox{\linewidth}{!}{\includegraphics{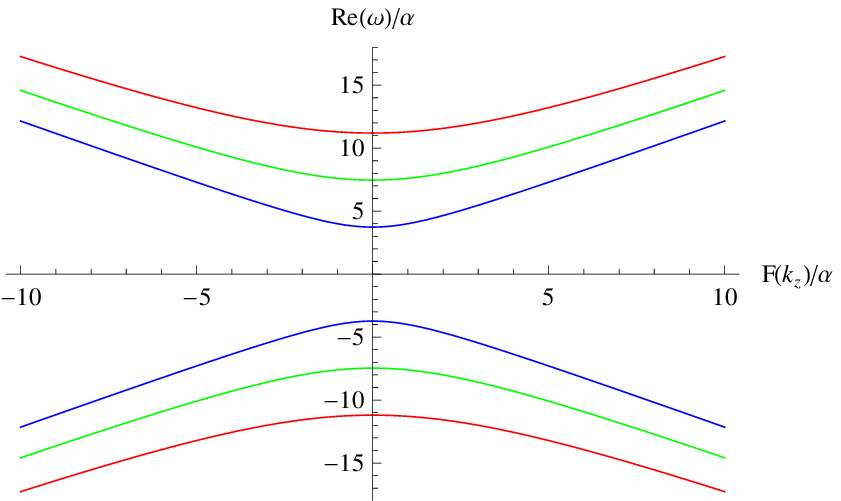}\includegraphics{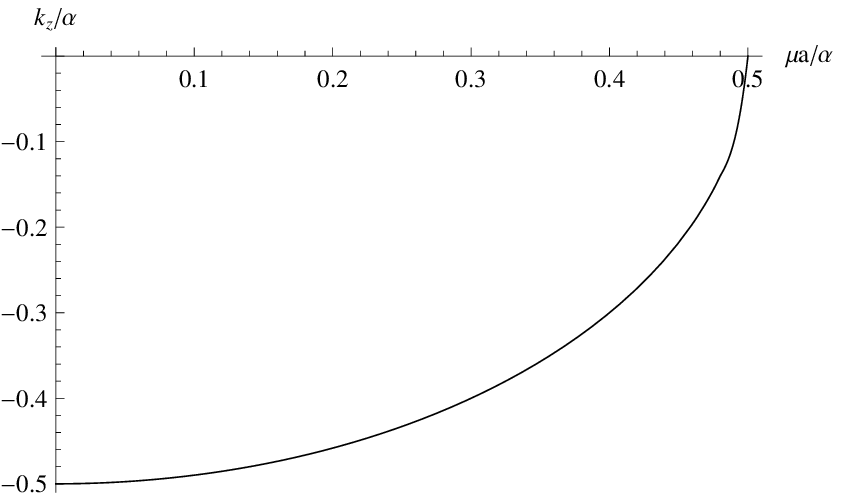}}
\caption{Left panel: real parts of Dirac quasinormal modes as functions of $F(k_z)$ for $B=1/2$. Upper (positive real part) curves correspond to $m=-1$, lower (negative real part) correspond to $m=+1$. Right panel shows $k_z$ as a function of $\mu$ for which $F(k_z)$ is zero.}\label{DiracQNMs}
\end{figure*}
\begin{figure*}
\resizebox{\linewidth}{!}{\includegraphics{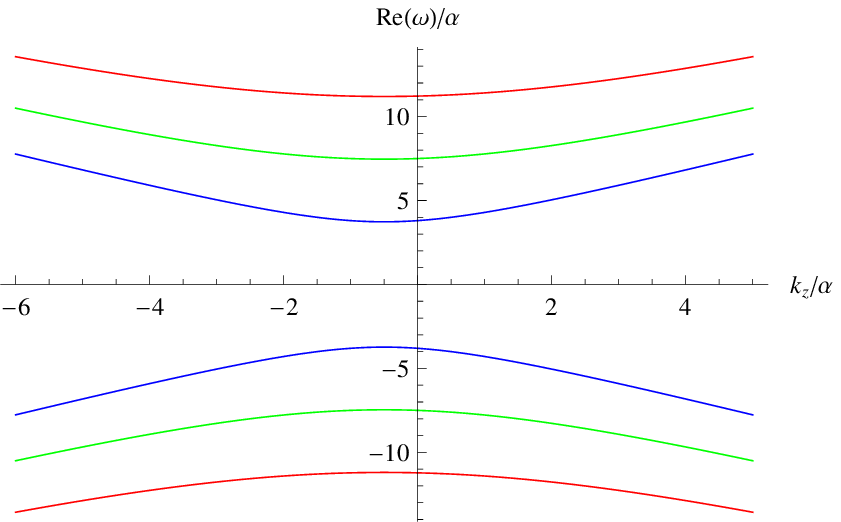}\includegraphics{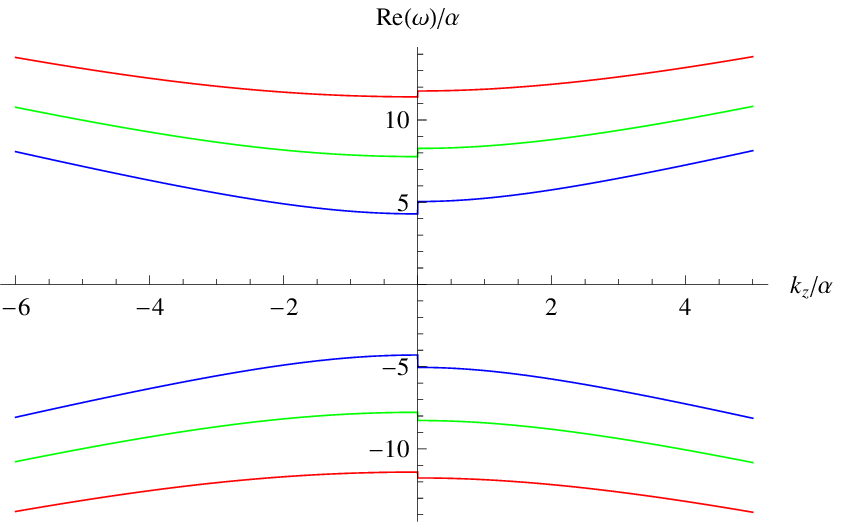}}
\caption{Real parts of Dirac quasinormal modes as functions of $k_z$ for $B=1/2$, $\mu=0$ (left panel) and $a\mu=2$ (right panel).}\label{DiracQNMsM}
\end{figure*}

One can see, that for $H\neq0$ the frequencies have an imaginary part of order $aH$, which corresponds to the decay of the amplitude due to expansion of the Universe. The real part of the frequencies form qualitatively the same picture as that for $H=0$ (see Figs. \ref{DiracQNMs},\ref{DiracQNMsM}).

From Fig. \ref{DiracQNMsM} we observe that quasinormal spectra are not the same for particles of opposite chiralities. The spectrum for the right-handed chirality can be obtained by the replacing $k_z\rightarrow-k_z$ in (\ref{left-handed}). Since for massless particles the helicity is an invariant quantity, we observe a continuous dependence of the QNMs on $k_z$ (left panel of Fig. \ref{DiracQNMsM}). Discontinuity for the massive particles of the same chirality as $k_z\rightarrow 0$ (right panel of Fig. \ref{DiracQNMsM}) reflects a nontrivial dependence of QNMs on helicity, which is not an invariant quantity in this case. The left and right limits as $k_z \rightarrow 0$ correspond to particles with different dominant helicities that are described by the same left-handed chirality. If one considers opposite chiralities for the positive and negative values of $k_z$, implying domination of the same helicity, the figure is symmetric for both massive and massless particles and has no discontinuity when $\mu^\textbf{2}>0$.

Note, that normalizable solutions exist only when $k_y\neq0$. For $k_y=0$, Eq.~(\ref{scalar-wavelike}) reads
\begin{equation}
N_\omega''(\eta)=\frac{K}{\eta^2}N_\omega(\eta),
\end{equation}
whose general solution is
$$N_\omega(\eta)=\sqrt{\eta}(C_+\eta^p+C_-\eta^{-p}),$$
where $p^2=K+\frac{1}{4}$.

The only nondivergent solutions is constant when $p=\pm\frac{1}{2}$ and $K=0$ \cite{Leahy:1982dj}. However, for these solutions the integrals (\ref{norma}) and (\ref{normaDirac}) diverge.

\vspace{3mm}
\section{Conclusions}\label{sec:conclusions}

In this work we have shown how the cosmological rotation and expansion can influence the stability of tachyons and the spectrum of
classical tardyon fields.
The stability gap for tachyons was expected from the similarities between the spectrum of asymptotically G\"odel \cite{Konoplya:2011ig,Konoplya:2011it,Konoplya:2005sy} and AdS space-times. Indeed, in  \cite{Konoplya:2011ig,Konoplya:2011it} it was noted that the spectrum of black holes in the 5-dimensional asymptotically  G\"odel space-time, is qualitatively similar to the normal modes of pure G\"odel space-time and is quite different from the quasinormal modes of the black hole in asymptotically flat space-time. In the limit of the vanishing black hole radius, quasinormal modes approach the normal modes of the pure G\"odel space. The same behavior takes place for black holes in AdS space-time \cite{Konoplya:2002zu,Konoplya:2011qq}, where the anti-de Sitter radius $R$ plays the role of the Universe's rotation rate $\Omega$.

Here we have found that the spectrum and stability of Dirac tachyons is quite different from the scalar one: while the scalar tachyons do not need expansion for the stability gap, that is, the gap exists for rotating and not expanding models, Dirac tachyons are unstable in the nonexpanding cosmological models. Yet, for both scalars and neutrino tachyons the coupling of rotation and expansion can enlarge the stability gap, making thereby cosmological rotation an important stabilizing factor.

Finally, since we have shown that the upper bound on the tachyon mass in all of the above cases is quite small, and is definitely many orders less than the electron's mass, one could suppose that this could explan why today only the lightest particles (though with nonzero mass, such as neutrinos) could be tachyonic.

An important issue which was beyond our consideration is quantization of the considered fields in the expanding G\"odel-like backgrounds.
One can show that the spectrum of bound states considered here do not form a complete set, so that the quantization of massive fields in such space-times is a nontrivial question, not to mention the tachyonic nature of the fields. Yet, if such a quantization is performed, the effort will pay off, allowing for a quantum description of stable tachyons.

In addition, we have described the spectra of tardyon fields in detail. The spectrum of the massive scalar or Dirac fields do not form a complete set and do not have unstable normalizable modes. Frequencies for a massive Dirac field (as well as a massless one) are not symmetric relative to the $k_z$ axis, where $k_z$ is the component of the wave vector. A new feature which that we observe for the massive Dirac field is the discontinuity of the frequency as a function of $k_z$ at $k_z =0$.

\section*{Acknowledgments}
R.~K. acknowledges the European Commission for financial support through a Marie Curie International Incoming Contract.
A.~Z. was supported by Conselho Nacional de Desenvolvimento Cient\'ifico e Tecnol\'ogico (CNPq).

\end{document}